\documentclass[twoside]{article}
\usepackage{qic,epsfig}

\begin{document}
\setlength{\textheight}{8.0truein}    

\runninghead{Precise creation, characterization, and manipulation of
single optical qubits}
            {N. Peters, J. Altepeter, E. Jeffrey, 
D. Branning and P. Kwiat}

\normalsize\textlineskip
\thispagestyle{empty}
\setcounter{page}{1}


\vspace*{0.88truein}

\alphfootnote

\fpage{503}

\centerline{\bf
Precise Creation, Characterization, and Manipulation of Single Optical Qubits
}
\vspace*{0.035truein}
\centerline{\bf 
}
\vspace*{0.37truein}

\centerline{\footnotesize 
Nicholas Peters, Joseph Altepeter, Evan Jeffrey, 
David Branning, and Paul Kwiat
}

\vspace*{0.015truein}
\centerline{\footnotesize\it Department of Physics, 
University of Illinois, 1110 W. Green St.}
\baselineskip=10pt
\centerline{\footnotesize\it Urbana, Illinois 61801,USA}
\vspace*{10pt}

\vspace*{0.225truein} \publisher{June 28, 2003}{September 16, 2003}

\vspace*{0.21truein}

\abstracts{
We present the theoretical basis for and experimental verification of arbitrary single-qubit 
state generation, using the polarization of photons generated via spontaneous
parametric downconversion.  Our precision measurement and state reconstruction system has 
the capability to distinguish over 3 million states, all of which can be reproducibly 
generated using our state creation apparatus.  In order to complete the triumvirate of single qubit 
control, there must be a way to not only manipulate single qubits after creation
and before measurement, but a way to characterize the manipulations
\emph{themselves}.  We present a general representation of arbitrary processes,
and experimental techniques for generating a variety of single qubit
manipulations, including unitary, decohering, and (partially) polarizing
operations.
}{}{}

\vspace*{10pt}

\vspace*{3pt}
\communicate{J Shapiro \& H-K Lo}

\vspace*{1pt}\textlineskip  
\section{Introduction}
Quantum information processing (QIP) promises great potential power over its
classical counterpart in the areas of computing, communication, and
metrology~\cite{nc}.  Nearly all QIP protocols require specific initial 
states and the ability to manipulate the qubits with exquisite
precision.  Here we demonstrate the creation of arbitrary single qubit states 
encoded in the polarization of single photons.  Single-photon Fock states are
conditionally realized by detecting one photon of a pair produced in the
process of spontaneous parametric downconversion~\cite{hong}.  We manipulate
the photons' polarization state using a series of birefringent waveplates
(which enable any unitary transformation) and a thick birefringent decoherer
(which allows the production of mixed states). Using a method of state
tomography, we can experimentally determine the most likely density matrix
which describes the resulting quantum state.  Finally, we have implemented
several versions of {\it process} tomography, by which we can accurately
characterize any quantum process acting on the polarization qubit, including
all unitary transformations, (partial) measurements, and decoherence.  

\section{State Creation}

We represent our qubits by single-photon polarization states in the
horizontal-vertical basis, with the state vectors
\begin{equation}
|0\rangle \equiv |H\rangle \equiv \left( \begin{array}{c} 1 \\
 0  \end{array}\right )~{\rm and}~ 
|1\rangle \equiv |V\rangle \equiv \left( \begin{array}{c} 0 \\
 1  \end{array}\right ), 
\end{equation}
or in density matrix notation, 
\begin{equation}
\rho_H = \left( \begin{array}{cc}1 & 0 \\
0 & 0 \end{array} \right )~{\rm and}~ \rho_V = \left( \begin{array}{cc}0 &
0 \\
0 & 1 \end{array} \right ). 
\label{rhoH} 
\end{equation}
The density matrix of an arbitrary single qubit can be represented by
three independent real parameters ($A$, $B$, and $\delta$):
\begin{equation}
\rho = \left( \begin{array}{cc}A & B e^{i \delta} \\
 B e^{ -i \delta} & 1-A \end{array} \right ), 
\label{arbstateeq} 
\end{equation} 
where 0$\leq$ A $\leq$1 from normalization, and $|B|\leq \sqrt{A(1-A)}$ from
positive semi-definiteness.  Another equivalent representation is given by
\begin{equation}
\rho= \frac{1}{2}( {\bf I}+{\bf \vec{r}}\cdot \mathbf{\sigma}),
\label{sigmastateeq} 
\end{equation}
where $\bf{I}$ is the identity and we define the polarization analogs of
the Pauli spin matrices as
\begin{equation}
\sigma_1 \equiv \left( \begin{array}{cc}0 & 1 \\
1 & 0 \end{array} \right ),~
\sigma_2 \equiv \left( \begin{array}{cc}0 & -i \\
 i & 0 \end{array} \right ),~{\rm and~} \\
\sigma_3 \equiv \left( \begin{array}{cc} 1 & 0 \\
0 & -1  \end{array} \right ). 
\label{spinmatrices}
\end{equation} 

The components of ${\bf \vec{r}}$ give the degree of polarization for the
photon in the Horizontal-Vertical (H-V), Diagonal-Antidiagonal (D-A), and
Right-Left Circular (R-L) bases\footnote{These components $r_i$ are related 
to the Stokes parameters ($S_i$) of classical optics~\cite{bw} by 
$r_i=\frac{S_i}{S_0}$.}. As such, they are often identified as $r_H \equiv r_1$,
$r_D \equiv r_2$, $r_R \equiv r_3$\footnote{The
conversion between the representations in \ref{arbstateeq} and 
\ref{sigmastateeq} is given by 
$r_H  =  2A-1$, $r_D = 2B\cos(\delta)$, and $r_R = 2B\sin(\delta)$.}. 
It is useful to view these components as coordinates in a 3-D space of 
polarizations; the constraint $|{\bf \vec{r}}| \leq 1$ implies that all states must
lie inside or on a sphere of unit radius, known as the Poincar\'e sphere.
Points on the {\it surface} of the
sphere ($|{\bf \vec{r}}|=1$) represent pure polarization states (linear
polarization states on the equator, right and left circular polarization
on the north and south pole, respectively), while points inside the surface ($|{\bf \vec{r}}|<1$) represent partially mixed states.  The center of the sphere
($|{\bf \vec{r}}|=0$) corresponds to a completely mixed state, i.e., an
unpolarized photon.

Because an arbitrary state  has three independent 
parameters, the generation of such a state requires three adjustable elements.
Considering the Poincar\'e sphere, we make an ansatz that a half-waveplate
(HWP), followed by a thick birefringent decoherer, a half-waveplate,
and a quarter-waveplate (QWP) are sufficient to generate all one-qubit
polarization states from a pure linear polarization fiducial state
($|H \rangle$ in our case).  We now derive formulae that give waveplate
settings for an arbitrary state, thus proving our ansatz.  The operators
that represent the action of half- and quarter-waveplates, respectively,
are the Jones matrices~\cite{fowles}
\begin{equation}
\mathbf{O}_{HWP}(\theta) \equiv \left( \begin{array}{cc}-\cos{2\theta} &
-\sin{2\theta} \\
-\sin{2\theta} & \cos{2\theta} \end{array} \right ) 
\label{HWPop} 
\end{equation}
and
\begin{equation}
\mathbf{O}_{QWP}(\theta) \equiv \left( \begin{array}{cc}1-(1+i)\cos^2{\theta}
& -(1+i)\sin{\theta}\cos{\theta}  \\
-(1+i)\sin{\theta}\cos{\theta} & 1-(1+i)\sin^2{\theta} \end{array} \right
), 
\label{QWPop} 
\end{equation}
where in each case the parameter $\theta$ is the angle the optic axis
makes with horizontal.   

\begin{figure} [t!]
\begin{center}
\epsfig{file=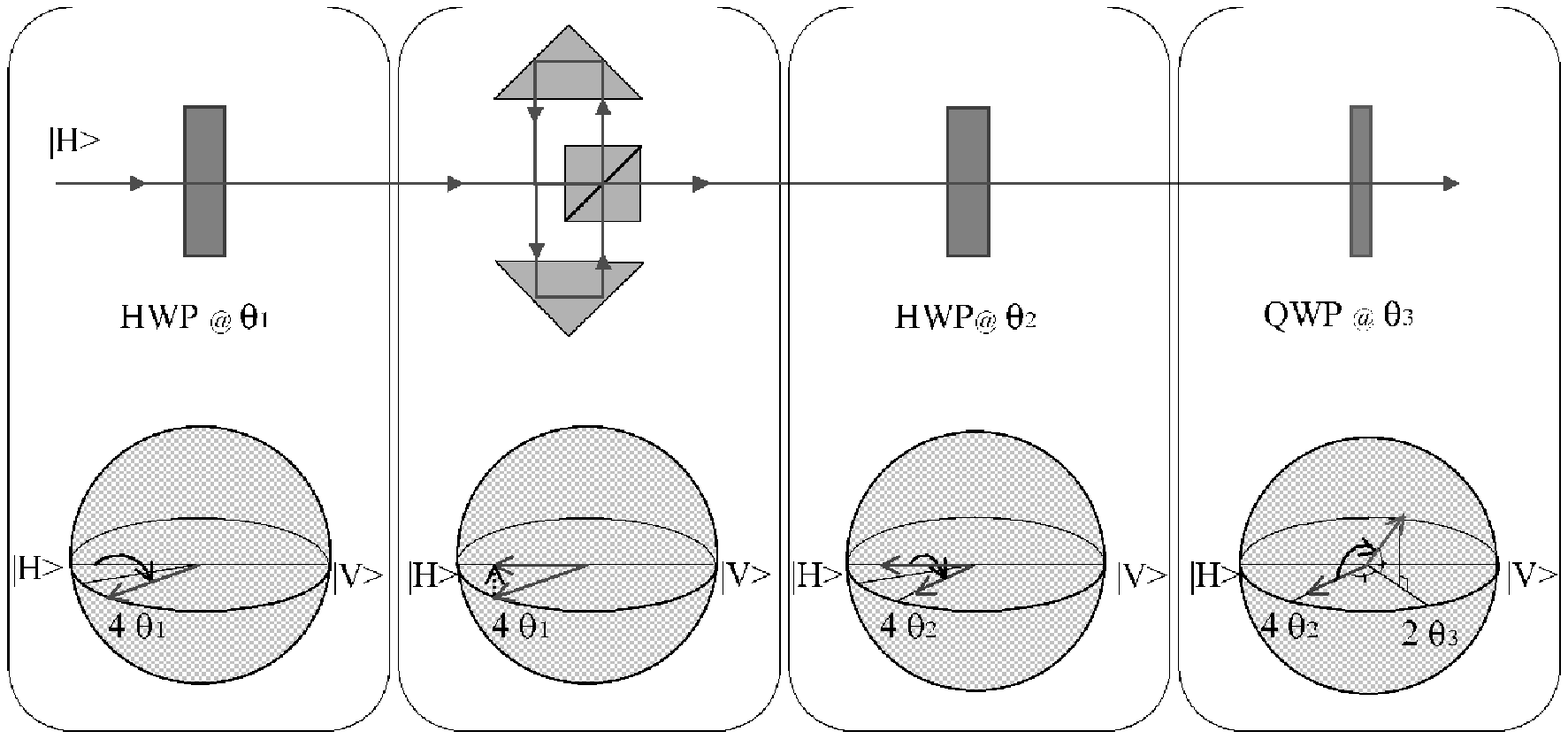, width=5truein}	
\vspace{5 mm}
\fcaption{The experimental setup to realize an arbitrary single (polarization) qubit, along with the representation of the state in the
Poincar\'e sphere at each step of state preparation.  In the first stage, on
the left, horizontally polarized photons are sent through a half-waveplate
(HWP) with an optic axis at $\theta_1$ from horizontal, giving the linear
polarization state $|2\theta_1\rangle$.  This is then sent through a
decoherer that has an optical path length difference greater than the
coherence length of the photon.  One method to achieve such a decoherer
is with a polarizing beamsplitter that sends vertical polarization in a
delay loop while horizontal is transmitted, thereby suppressing the phase
coherence in the horizontal-vertical basis (see Appendix A), and eliminating 
any amplitude
in the off-diagonal elements of the state.  Decoherence in the Poincar\'e
sphere appears as a projection of the state onto the $|H\rangle-|V\rangle$
``spindle'' (second box).  By appropriately adjusting the first HWP, states
with arbitrary mixedness can be produced, ranging from a pure state ($|{\bf
\vec{r}}|=1$) to a completely mixed state, i.e., unpolarized photons.
The last two waveplates, a HWP and a quarter-waveplate (QWP), set the
final direction of the (possibly mixed) state on the sphere.}
\label{figone}
\end{center}
\end{figure}

To create arbitrary states as in (\ref{arbstateeq}), we start with photons in
the state $\rho_H$, and direct them through a half-waveplate at $\theta_1$,
giving 
\begin{equation}
\mathbf{\rho_1} \equiv \mathbf{O}_{HWP}(\theta_1)\rho_H
\mathbf{O}_{HWP}(\theta_1)^{\dag}
= \left( \begin{array}{cc}\cos^2{2\theta_1} & \cos{2\theta_1}\sin{2\theta_1}
\\
\cos{2\theta_1}\sin{2\theta_1} & \sin^2{2\theta_1} \end{array} \right
)=|2\theta_1 \rangle \langle 2 \theta_1|, \label{rho1}
\end{equation}
i.e., the pure linear polarization state $|2\theta_1 \rangle$. As shown
in the first box in Fig.~\ref{figone}, this operation is described on the
Poincar\'e sphere by rotating the state $|H\rangle$ by $180^{\circ}$ about an 
axis -- representing the optic axis of the waveplate -- that lies $2\theta_1$
away on the equator; the factor of 2 arises because $|V\rangle$ lies on
the {\it opposite} side of the Poincar\'e sphere, i.e., $180^{\circ}$
away from $|H\rangle$\footnote{A very complete discussion of the use of the
Poincar\'e sphere to describe the action of various crystal optics may be found
in~\cite{ramachandran}.}.

The next step is to introduce decoherence by separating the
horizontal and vertical polarization components by an optical path length
difference much longer than the coherence length of the photons (see
Appendix A)~\cite{berglund}.  If the coherence length is longer than
a few millimeters, one can use a polarization-dependent delay line, as
shown in the second box of Fig.~\ref{figone}.  For our downconversion
source, interference filters at the detector typically define the
spectral bandwidth; for filters with 10-nm full width at half maximum,
the coherence length is only 49-$\mu$m.  In this case it suffices to
use a thick birefringent element [e.g., 1 cm of quartz] to completely
decohere the polarizations within the eigenbasis of the decohering 
element~\cite{berglund}.  We can control the {\it amount} of decoherence by
tuning the polarization of the input state. For example, if the polarization
before entering the decoherer is $|H\rangle$ (or $|V\rangle$) then the
state purity is preserved; if the input state is diagonally polarized, i.e.,
$|D\rangle$, the resulting state
is completely mixed.  An arbitrary value of $|{\bf \vec{r}}|$ is produced by 
setting the
orientation of the first half-waveplate to $\theta_1=\frac{1}{4}\arccos{|{\bf
\vec{r}}|}$.  After the rotated light is directed through the birefringent
decohering element, the reduced density matrix describing only the polarization
is of the form (see Appendix A): 
\begin{equation} \mathbf{\rho_1'} = \left( \begin{array}{cc}\cos^2{2\theta_1}
& 0 \\
0 & \sin^2{2\theta_1} \end{array} \right ). \label{rho1prime} \end{equation}
Next, using waveplates, we unitarily transform (\ref{rho1prime}) into
the desired final state.  Note that (\ref{rho1prime}) can be rewritten as
an incoherent sum of a horizontally polarized pure state and an unpolarized,
completely mixed state:  
\begin{equation}
\mathbf{\rho_1'} = \cos 4\theta_1|H\rangle \langle H|+2\sin^2 2\theta_1
\rho_{mixed}\,,
\end{equation}
where $\rho_{mixed} \equiv \frac{1}{2} {\bf{I}}$ is the completely mixed state.
Because quantum mechanics is linear, we may operate individually on each
part of this sum to realize the final state.  As the unpolarized part is is unchanged by any unitary
transformation, the final form of the state is determined by operating
on the $|H\rangle \langle H|$ term with a half-waveplate (at $\theta_2$)
and a quarter-waveplate (at $\theta_3$).  Algebraically determining the
desired values of $\theta_2$ and $\theta_3$ to obtain a particular final
state is non-trivial.  However, by noting the geometric action of these
waveplates on the Poincar\'e sphere (see the third and fourth boxes of
Figure~\ref{figone}), one can arrive at the following final solutions for the waveplate angles\footnote{In terms of the Poincar\'e sphere
vector ${\bf \vec{r}}$, we have \begin{eqnarray} \nonumber \theta_1 & =
& \frac{1}{4} \arccos{(|{\bf \vec{r}}|)} \\ 
\nonumber\theta_2 & = & {1 \over
4}\left[\arctan{(\frac{r_D}{r_H})}+\arctan{(\frac{r_R}{\sqrt{r_H^2+r_D^2}}})\right]
\\ \nonumber \theta _3 & = &  \frac{1}{2}\arctan{(\frac{r_D}{r_H})}.
\end{eqnarray}} 
\footnote{In any real system, the various waveplates may have slight errors in
their retardations.  Nevertheless, we have shown that any single qubit state
may be produced, as long as the QWP retardance error is 
smaller than the HWP retardance error.  In this case, however, we must numerically search for the
optimal waveplate settings.}: 
\begin{eqnarray}
\theta_1 & = & \frac{1}{4} \arccos [\sqrt{(2 A - 1)^2 + 4 B^2}] \\ 
\theta_2 & = & {1 \over 4} \left[\arctan\left[ 2 B \cos [ \delta ]\over
2A-1 \right] + \arctan\left[ 2 B \sin [ \delta ] \over \sqrt{(2 A - 1)^2 +
4 B^2\cos^2 [\delta] } \right]\right] \\
 \theta _3 & =&  \frac{1}{2}\arctan\left[ 2 B \cos[ \delta ]\over
 2A-1\right].
\end{eqnarray}

\begin{figure}
\begin{center}
\epsfig{file=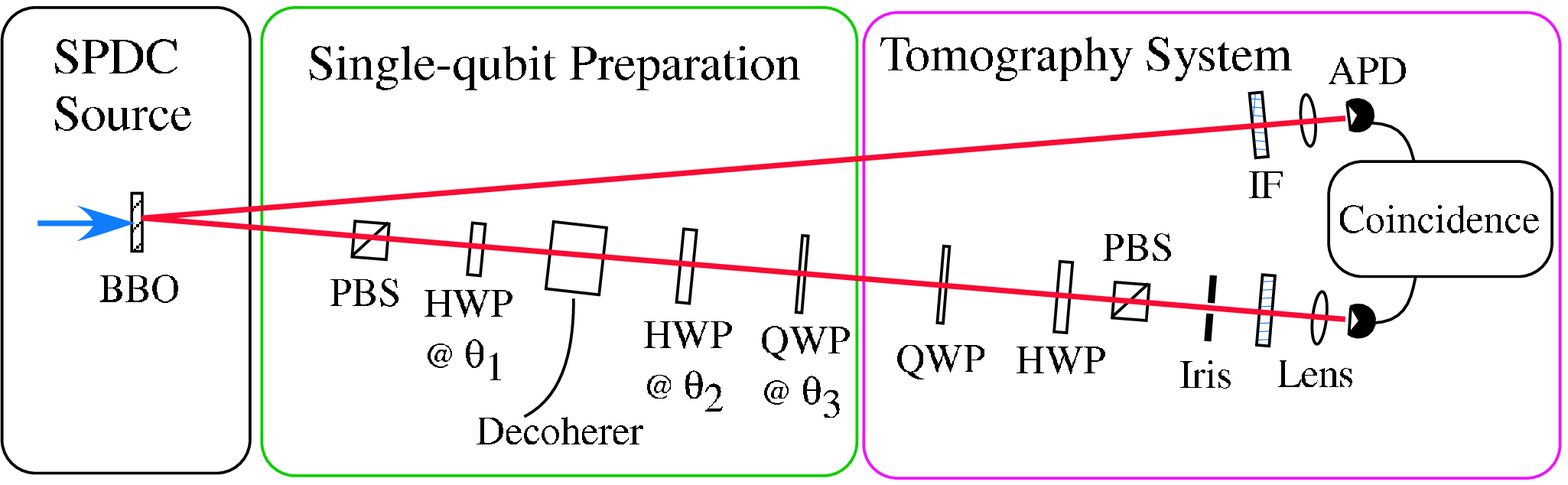, width=5truein}
\vspace{5 mm}
\fcaption{Single-qubit downconversion experiment. Detection of a photon in
the upper arm conditionally prepares a single-photon state in the lower arm.
The polarizing beamsplitter (PBS) after the nonlinear BBO crystal prepares
an extremely pure initial state $|H\rangle$, which is then transformed by
the waveplates and decoherer ($\sim$1 cm quartz birefringent element).
The tomography system allows an accurate measurement of the resulting
density matrix $\rho$ (see Sect. 3).}
\label{qubitex}
\end{center}
\end{figure}

Armed with these equations we set up our experiment as in Figure~\ref{qubitex}.
The process of spontaneous parametric downconversion 
conditionally prepares single-photon input states~\cite{hong}.  A nonlinear crystal
(BBO) is pumped with a vertically polarized 351-nm beam from an Argon-ion
laser (average power 86 mW).  The BBO is cut to produce non-collinear
frequency-degenerate horizontally polarized photon pairs at 702 nm via
type-I phase matching.	The non-collinear photon pairs are collected
in two modes, separated by $6^{\circ}$ outside of the crystal.  
The first mode impinges onto a
detector assembly consisting of a 10-nm FWHM interference filter (IF)
centered at 702.2 nm, a lens, and an avalanche photodiode (APD) operated
in Geiger mode (Perkin-Elmer \#SPCM-AQR-14).  Single-photon Fock states are
prepared in the second arm conditional on a ``trigger'' count in the first
detector~\cite{hong}.  The second mode passes through a polarizing beamsplitter
(PBS) (to ensure a good fiducial state $|H\rangle$), the state preparation
waveplates\footnote{All of the waveplates are in motor-controlled
stages, and can be set with an accuracy of less than $0.1^{\circ}$.  }~~and
decoherer (1 cm quartz slab) discussed above, a QWP-HWP-PBS combination
for state analysis (see Sect. 3), a 2.2-mm iris, and a detector assembly
identical to that in the first mode. 

\begin{figure}
\begin{center}
\epsfig{file=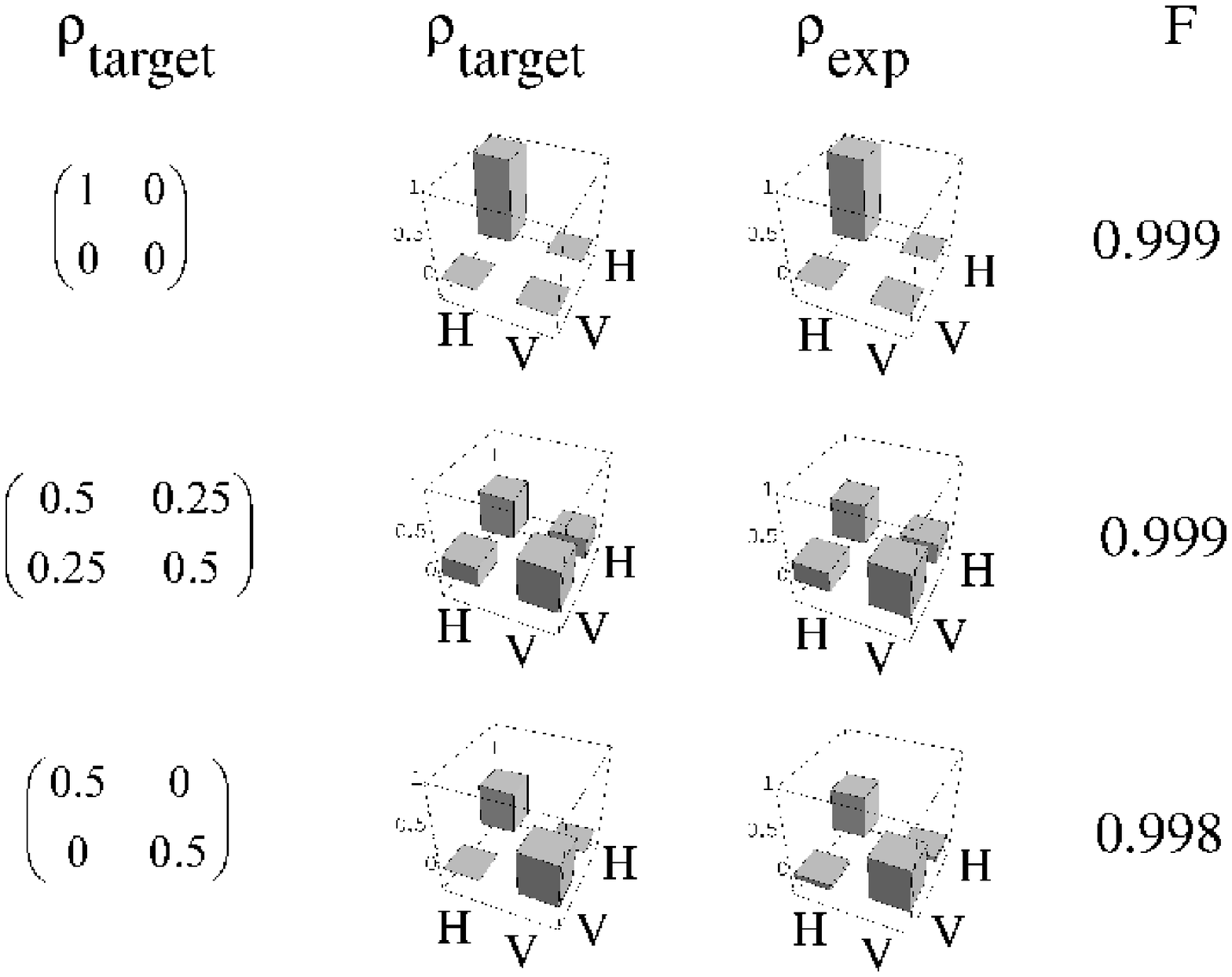, width=3.5truein}	
\fcaption{Single qubit data.  Shown are the target density matrices, plots of the target and experimentally measured density matrices, and the fidelity (F) between them.
The imaginary elements of the density matrices are not shown because they
have zero amplitude for the target states, and are always less than 2\%
for the experimental states.}
\label{figdata}
\end{center}
\end{figure}

We created a variety of single qubit states using this system.  Some sample
reconstructed density matrices are shown in Figure~\ref{figdata}.
One measure of our ability to accurately prepare specific states
is the fidelity\footnote{The fidelity is a measure of state
overlap~\cite{jozsa}.  For two pure states, $|\psi_1\rangle$ and
$|\psi_2\rangle$, the fidelity is simply 
$F(|\psi_1\rangle,|\psi_2\rangle)\equiv|\langle
\psi_1|\psi_2\rangle|^2$, while for two general
density matrices, $\rho_1$ and $\rho_2$, the fidelity is
$F(\rho_1,\rho_2)\equiv|Tr(\sqrt{\sqrt{\rho_1}\rho_2\sqrt{\rho_1}})|^2$.}~~between the measured and the target states.  
We typically observed fidelities better than 0.997; this number is presently
limited by counting fluctuations due to Poisson statistics,
and also to small drifts (less than 0.5\%) in either the pump laser 
intensity or the detector efficiencies. For the data in Fig.~\ref{figdata}, a 
total $\sim$150,000 counts were accumulated for each state.

\section{State Characterization}
One cannot determine the unknown polarization state of a photon with a
single measurement.  Instead, a large ensemble of photons prepared in
an identical manner must be projected into different polarization basis
states.  Four measurements are needed -- three to determine the
relative values of the three independent parameters that characterize an
unknown state, and a fourth to determine normalization.  The polarization
analysis is carried out setting the quarter- and half-waveplates shown in
the tomography system box of Fig.~\ref{qubitex}.  Although there are many choices
as to the particular measurements that can be made (in principle measuring
along any three non-collinear bases suffices), we choose the analysis
states: $\langle \psi_0| \equiv \langle V|$, $\langle \psi_1|\equiv \langle H|$,
$\langle \psi_2| \equiv \langle D|$, and
$\langle \psi_3| \equiv \langle R |$
corresponding to $N_0$, $N_1$, $N_2$, and $N_3$ photon counts in some
fixed measurement time interval (typically 100 s).  As explained earlier, 
to obtain single
photon Fock states, we count in coincidence; thus, $N_i$ are coincidence
counts between the two detectors shown in Fig.~\ref{qubitex}.  Since we
need to estimate probabilities, we must measure a complete basis to
normalize the photon counts.  As $|H\rangle$ and $|V\rangle$ form a basis,
${\cal N}\equiv N_0+N_1$ gives a convenient normalization factor for the $r_i$:
$r_H=2N_1/{\cal N}-1$, $r_D=2N_2/{\cal N}-1$, and $r_R=2N_3/{\cal N}-1$.
The density matrix of the state may then be reconstructed as in
(\ref{sigmastateeq}).

Unfortunately, as this state reconstruction is based on photon counting,
statistical fluctuations or drift often yield an unphysical result;
therefore, we employ a maximum likelihood technique to estimate a physical
density matrix that would most likely produce the measured data.  James et
al. describe the technique for determining the joint state
of {\it two} qubits~\cite{james}.  Here we distill their argument down
to the one qubit case (if states other than single-photon states are
used, more sophisticated methods must be used to determine the
quantum polarization state~\cite{raymer}).  

A physical density matrix representation has three cardinal
properties: normalization, positive semidefiniteness, and hermiticity.
Therefore, we first guess a density matrix that by definition
has the aforementioned properties.  A matrix that can be written
as $\mathbf{T}^{\dagger}\mathbf{T}$ is positive semidefinite and
hermitian~\cite{james}.  To normalize such a matrix, we divide by its trace
so that $\mathbf{T}^{\dagger}\mathbf{T}/Tr(\mathbf{T}^{\dagger}\mathbf{T})$
has the three properties for a legitimate physical density matrix.
We choose the following invertible form for $\mathbf{T(t)}$: 
\begin{equation}
\mathbf{T}(\mathbf{t})\equiv \mathbf{T}(t_1,t_2,t_3,t_4)\equiv \left(
\begin{array}{cc} t_1 & 0 \\
t_3+i t_4 & t_2 \end{array} \right ). \label{Top} \end{equation}
Using (\ref{Top}), the density matrix formula is
\begin{equation}
\mathbf{\rho}_p(t_1,t_2,t_3,t_4) =
\mathbf{T}^{\dagger}(\mathbf{t})\mathbf{T}(\mathbf{t})/Tr(\mathbf{T}^{\dagger}(\mathbf{t})\mathbf{T}(\mathbf{t})).
\label{tfunction} \end{equation}
Next we will introduce a likelihood function that quantifies how similar
$\rho_p(t_1,t_2,t_3,t_4)$ is to our experimental data:
\begin{equation}
{\cal L}(N_0,N_1,N_2,N_3;t_1,t_2,t_3,t_4)= \sum_{\nu=0}^3\frac{[{\cal N}
\langle \psi_\nu|\rho_p(t_1,t_2,t_3,t_4)| \psi_\nu \rangle-N_\nu ]^2}{{2
\cal N} \langle \psi_\nu|\rho_p(t_1,t_2,t_3,t_4)| \psi_\nu \rangle},
\end{equation} 
where the quantity ${\cal N} \langle \psi_\nu|\rho_p(t_1,t_2,t_3,t_4)|
\psi_\nu \rangle$ represents the expected number of counts for a projection
of our trial density matrix $\rho_p$ onto the analysis state $|\psi_\nu
\rangle$.  Note that the coincidence counts are subtracted from ${\cal
N} \langle \psi_\nu|\mathbf{\rho}_p(t_1,t_2,t_3,t_4)| \psi_\nu \rangle$
so that the likelihood function must be {\it minimized} to obtain the set of
$t_i^{(opt)}$, and therefore the state that best retrodicts the actual 
measurements\footnote{The minimization is performed using the {\sf MATLAB R12}
function {\sf fminunc} which requires an initial estimate for the
$t_i$.	For the initial estimate, starting guesses for the $\rho_i$ are
calculated from the measured counts $N_0$, $N_1$, $N_2$, and $N_3$; through
(\ref{sigmastateeq}), this gives a starting estimate for $\rho$, which, from
(\ref{Top}) and (\ref{tfunction}), yields starting values for the $t_i$:
$t_1  =  1 -\sqrt{1-\frac{1}{2(1-r_H)}((1-r_H)^2+r_D^2 + r_R^2)}$,
$t_2  =  \frac{ \sqrt{2(1-r_H)}}{2}$,
$t_3  = \frac{r_D}{\sqrt{2(1-r_H)}}$, and
$t_4  =  \frac{r_R}{\sqrt{2(1-r_H)}}$.}.

\begin{figure}
\begin{center}
\epsfig{file=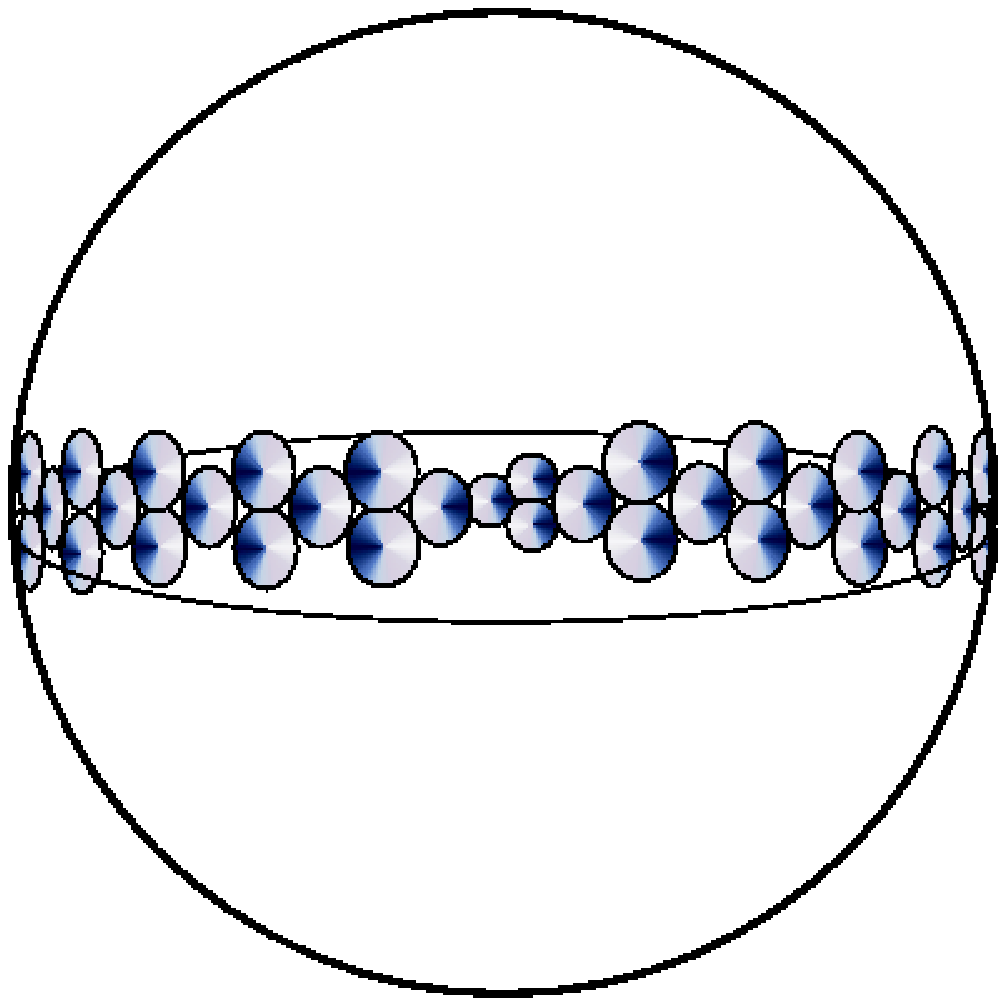, width=4cm}
\vspace{5 mm}
\fcaption{Filling the Poincar\'e sphere with experimentally determined
uncertainty ``patches".  These arise primarily due to counting
fluctuations, either from the Poisson statistical uncertainties in photon
counting, or from slight drifts ($<$ 0.5\%) in the laser power or detector
efficiencies.  The measured uncertainty patches are approximately  ellipsoids
(pancakes) as shown above; for clarity their dimensions are shown scaled up by
a factor of 5.  Note that the thickness of the pancakes depends somewhat
on the mixture of the state, indicating that some regions of the Poincar\'e sphere are
more sensitive to counting statistics.  The size of uncertainty patches
implies that we are able to distinguish more than three million unique 
single-qubit states, assuming $\sim$10 minutes
data collection time per state, i.e., $\sim$300,000 detection events).}
\label{PSphere}
\end{center}
\end{figure}

The question of how many different states can be reliably {\it
produced} is limited by how many may be reliably experimentally {\it
distinguished}\footnote{When specifying a produced quantum state, it
is an interesting question whether one should give an error ball (e.g., a
patch on the surface of the sphere for a pure state), or simply average over
this ball to yield a slightly mixed state.  The former approach accounts
for the fact that the error ball could shrink if more data were taken.
The latter method acknowledges that the density matrix already encodes
the totality of our knowledge, based on our measurements on identically
prepared members of an ensemble.}.  To determine the number of accessible
distinguishable states in our system, several states in different parts of
the Poincar\'e sphere (with approximate $|{\bf \vec{r}}|$ values of 0, 0.25,
0.5, 0.75, and 1) were created and measured 10 times each.	We measured
for 10 minutes per tomography yielding 300,000
counts per state reconstruction.  For each set of measurements the average
state vector $\hat{{\bf \vec{r}}}$ was determined.  Next, the standard deviation 
of the ten trials was determined for three directions: 
$\hat{{\bf \vec{r}}}$, and two directions transverse to $\hat{{\bf \vec{r}}}$.
Assuming a Gaussian distribution along these directions, we calculate that an
ellipsoid with semi-axes equal to 1.69 times these standard deviations will 
account for over 95\% of the events. These uncertainty ellipses are mainly due
to count fluctuations from laser and detector efficiency drift, as well as 
intrinsic Poissonian counting statistics.  
One unexpected result of our measurements was that the thickness of the
ellipsoid, i.e., the length of the minor axis along the direction of $\hat{{\bf \vec{r}}}$,
depended on the radial coordinate $|\hat{{\bf \vec{r}}}|$, and varied from a 
minimum value of 0.0021 for $|\hat{{\bf \vec{r}}}|=1$, to a maximum value of 
0.0062 for $|\hat{{\bf \vec{r}}}|=0.25$ (see Figure~\ref{PSphere}).  
Numerical simulations support this observation.
Taking into account the varying size of the uncertainty patches, and assuming an approximate close packing of the entire Poincar\'e sphere volume, we estimate that we can reliably distinguish over three million states.

\section{State Manipulation}
\begin{figure}
\begin{center}
\epsfig{file=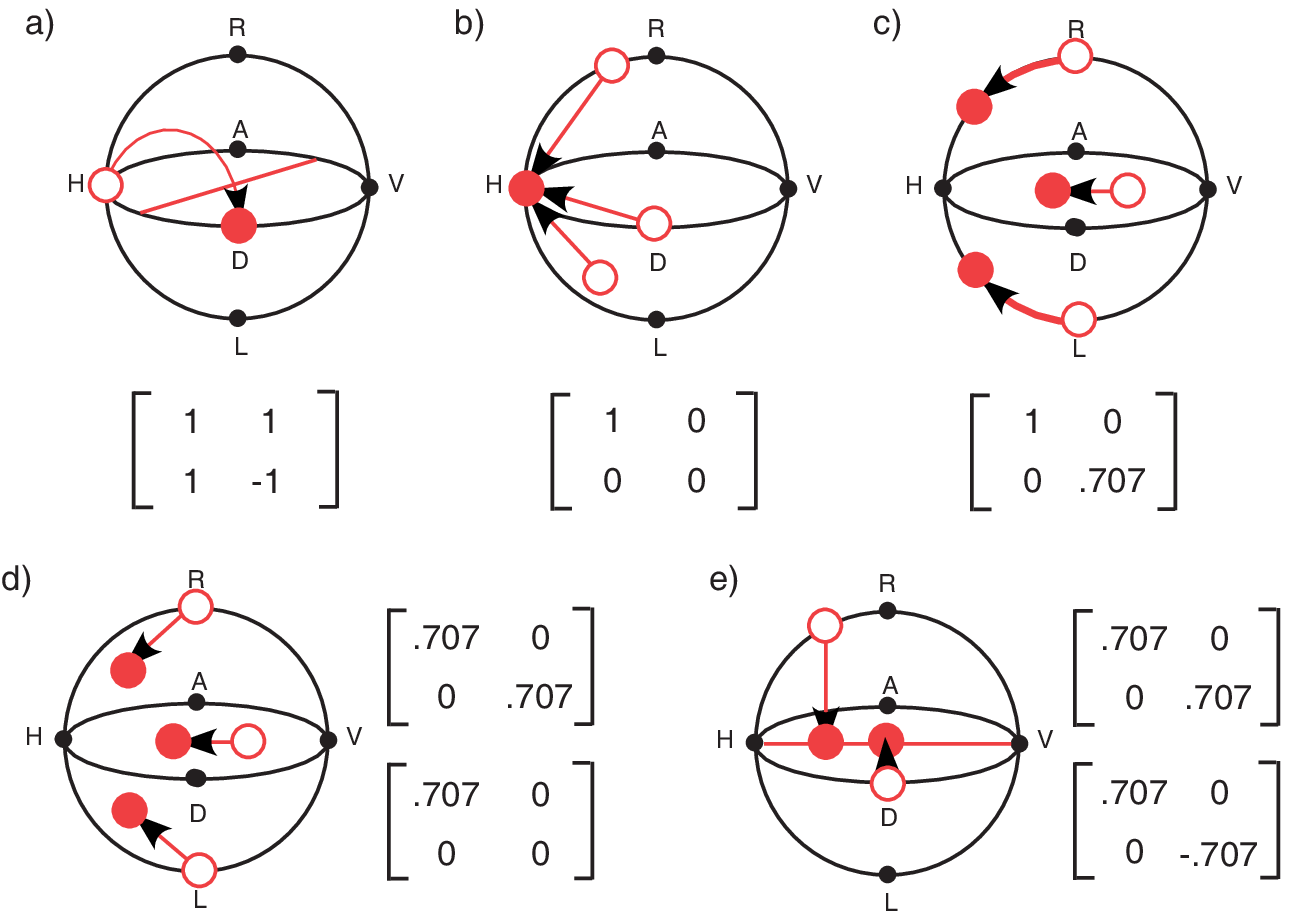, width=14cm}
\vspace{5 mm}
\fcaption{
Shown are several example single-qubit quantum processes.  Each process is represented
by both a geometric picture and by operator matrices.  The geometric
picture shows initial states (white balls) on the Poincar\'e sphere mapped to
final states (solid balls).  The effect of a process $\cal{E}$ on an input state
can be defined as 
${\cal E}(\rho_{in}) = \sum_j E_j \rho_{in} E_j^\dagger $.
The operator matrices shown correspond to the $E_j$
matrices.  
(a) An example of a unitary transform.  This transform, the Hadamard gate, 
corresponds to a $180^\circ$ rotation about the $22.5^\circ$ axis.
(b) A total projection operator.  This corresponds to a perfect horizontal
polarizer.
(c) A coherent partial polarizer.  This partial projection corresponds to a
device which transmits all horizontal light and half of vertical light.  These
two components maintain their phase relationship.  Note that pure input states
remain pure while all states travel towards H on the sphere.
(d) An incoherent partial polarizer.  This corresponds to an operation which
50\% of the time projects into the H basis and 50\% of the time does nothing.
The photons which are subjected to the first operation are incoherently added to
those of the second operation; thus this process transforms pure input states
into mixed states.
(e) A decoherer.  This operation decoherers in the H-V basis, removing the phase
relationship between these two components.  All states travel in a straight line
to the spindle connecting the H and V states.
}
\label{J1}
\end{center}
\end{figure}

In order to gain total control over a single qubit, it is necessary to have an
understanding of what operations are physically possible.  
Any physical process acts as a map from all possible input states to a
transformed set of output states.  For example, the identity process maps every
possible input state to itself.  In other words, it has no effect on any input
states.  Other familiar processes include the unitary transformation, which when
viewed in the geometric picture of the Poincar\'e sphere, is simply a rotation
around a fixed axis (Fig.~\ref{J1}a);  
projections which either partially or completely
project a state into a single basis (a single point on the Poincar\'e
sphere, see Fig.~\ref{J1}b); or (partial) decoherence, which partially or
completely collapses all states to the ``spindle'' formed by the eigenstates
of the decohering interaction (Fig.~\ref{J1}e).

Note that while some processes, such as the identity, projection, or the
unitaries, can be represented by a single 2 by 2 matrix (which operates on a two
element column vector representing the qubit, as in Jones calculus), some more
complex processes such as decoherence or incoherent partial polarization
require a different method of characterization.  Consider a simple operation
(e.g., unitary or projecting) $E$ acting on a general state $\rho_{in}$. The
output state is given by
\begin{equation}
 \rho_{out} = E \rho_{in} E^\dagger. 
 \label{basic_process}
\end{equation}
A general operation ${\cal E}$ cannot be represented 
using a single $E$ matrix, but instead
can be described using the following operator-sum decomposition~\cite{nc}
\begin{equation}
\rho_{out} = {\cal E}(\rho_{in}) = \sum_j E_j \rho_{in} E_j^\dagger .
\label{sum_over_e}
\end{equation}
An arbitrary number of $E$ operators can be used to represent a process, and
as we saw above, decohering and partially polarizing processes 
require at least two matrices.  The necessity of multiple $E$ operators hints at the variety of
processes which must be possible, and the 
subtlety involved in representing an arbitrary process.  Consider, 
for example, the partial polarizers shown in 
figures \ref{J1}c and \ref{J1}d. Figure \ref{J1}c shows a coherent
partial polarizer, which acts as a partial projector.  Only a single $E$ matrix is
necessary to represent it.  Experimentally, this projector is implemented using
a series of tilted glass plates for which $T_V/T_H$ is 0.5\footnote{When the plates are tilted so that the photons are incident at Brewster's
angle $56^\circ$ \protect{\cite{bw}}, the p-polarization is completely 
transmitted while the s-polarization suffers a reflective loss of $15\%$ 
per air-glass interface.
}. All of the vertical light which passes through the glass plates maintains 
a definite phase relationship with all of the horizontal light which passes 
through.  For this reason we call this a \emph{coherent} partial polarizer.

Now consider a process which totally projects into the horizontal basis, but
only acts on 50\% of the measured qubits (photons).  The light passing
through the polarizer has no phase relationship with the light that does not,
and an incoherent mixture results.  We call this process an
\emph{incoherent} partial
polarizer (Fig.~\ref{J1}d).  In fact, any process which transforms a pure
state into a mixed state must be represented by having one $E$
operator acting on a certain percentage of qubits, and a separate operator or
operators acting on the remaining qubits.  Decoherence acts in exactly this way.
For example, consider the case in which the input qubits are 
randomly and incoherently subjected to either a $\sigma_z$ rotation (in
the H-V basis) or the identity matrix.  Combining the 
photons from the first process with those from the second, we see that any
coherence between the H and V bases has been destroyed, 
collapsing output states to the H-V spindle on the
Poincar\'e sphere.  This action is illustrated in Fig.~\ref{J1}e.

Using equation (\ref{sum_over_e}), we can simplify the general
representation of a process.  
First consider that the $E$ operators above can be
expanded into a linearly independent basis of 2x2 operators, such as the
identity matrix ($\sigma_0$) and the Pauli matrices ($\sigma_1, \sigma_2, 
\sigma_3$): 
\begin{equation}
E_j = \sum_{i=0}^3c_i\sigma_i .
\end{equation}
Substituting into equation (\ref{sum_over_e}) and combining terms,
\begin{equation}
{\cal E}(\rho_{in}) = \sum_{i,j=0}^3 \sigma_i \rho_{in} \sigma_j \chi_{ij}.
\end{equation}

The $\chi_{ij}$ matrix~\cite{qpt}, a 4x4 
positive Hermitian complex number matrix, shares many
similarities with a state density matrix $\rho$.  (For an earlier, alternate characterization of quantum processes, see \cite{qpt2}).  While a density matrix shows
the coherent and incoherent combinations of four 
orthogonal state vectors that make up a state, the $\chi$ matrix shows how an arbitrary process is
represented by coherent and incoherent combinations of four 2x2 operators (e.g., the $\sigma$ 
matrices).  In the case of the density matrix, a basis change
can yield the same state written in a different orthogonal four-element
basis.  
There is an exact analog to this orthogonality condition for the $\chi$
matrix, with orthogonality defined as $Tr(E_1E_2)=0$.  In fact, if the
$\chi$ matrix is diagonalized, it becomes clear that an arbitrary process can be
represented by the operator-sum decomposition above, with only four orthogonal
2x2 operators in the sum.  Note that in general these four operators will
\emph{not} simply be the Pauli matrices $\sigma_i$.

\begin{figure}[h]
\begin{center}
\epsfig{file=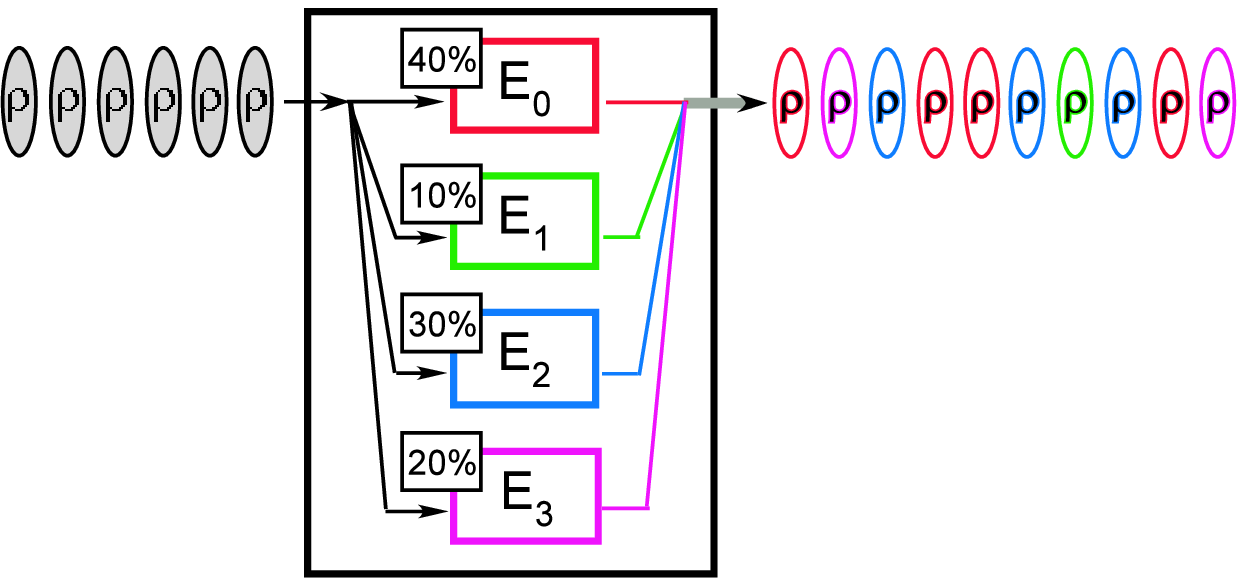, width=14cm}
\vspace{5 mm}
\fcaption{
An ensemble of single qubits are operated upon by an example of an arbitrary
process.  In general, any process can be represented by four orthogonal 2x2
operators, each of which has some percentage chance of affecting an input qubit.
These four operators have no restrictions, but they must in some way
incoherently act upon input qubits.  In this way input pure states affected
by more than one operator can be transformed into a mixed ensemble.
}
\label{J2}
\end{center}
\end{figure}

A physical picture of this representation corresponds to four different
orthogonal 2x2 operators, each of which has a specific \emph{probability} 
to act on an
input qubit, i.e., they are applied incoherently.  
In this way an ensemble of
identical input pure states can be transformed into a mixed output ensemble
(see Fig.~\ref{J2}).

\begin{figure}[h]
\begin{center}
\epsfig{file=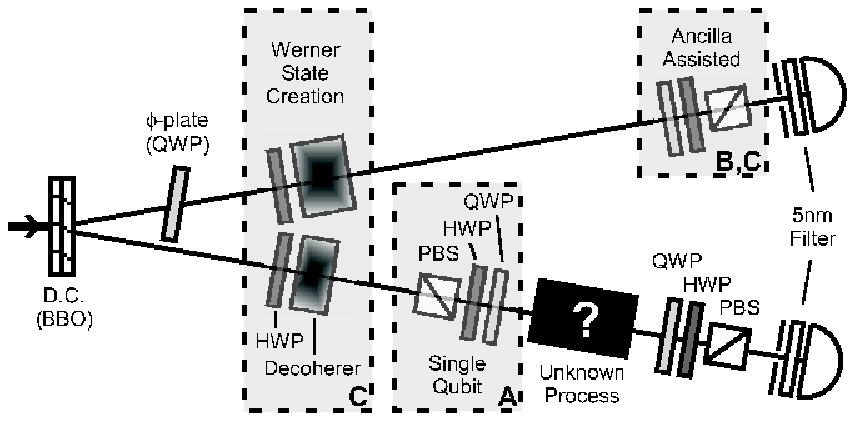, width=10.5cm}
\vspace{5 mm}
\fcaption{
Experimental setups for quantum process tomography. 
A, B, and C above denote which elements are present for SQPT, EAPT, and
unentangled AAPT, respectively. (a) Standard quantum process tomography:
Polarizer (P), half waveplate (HWP) and quarter
waveplate (QWP) allow preparation of required pure single-photon
states; identical elements
allow tomography of the post-process states. (b)
Entanglement-assisted process tomography: The source produces the
maximally entangled state $(|HH \rangle - |VV \rangle)/\sqrt{2}$.
A two-photon tomography of the output allows reconstruction of the
process. (c) Ancilla-assisted process tomography: 
The source produces a separable, or completely unentangled,
input state.  Correlations in this state still allow AAPT.
}
\label{J3}
\end{center}
\end{figure}

Now that we have a useful representation of arbitrary processes
and a physical interpretation of that representation, it is necessary to be able
to measure and reconstruct a given unknown process.  The measurement of an
unknown process is accomplished using quantum process tomography.
The simplest way to measure a process involves preparing a complete basis of
four single-qubit input states (e.g., H, V, D, R), subjecting each of these
to the unknown process, and measuring the four output states.  This is
referred to as standard quantum process tomography (SQPT)~\cite{qpt, qpt2}. 
While this is effective, it requires four input states.  It is possible, 
however, with a single \emph{2-qubit} input state to exactly characterize an
unknown process~\cite{D'Ariano, D'Ariano2, D'Ariano3}.  By using a second ancillary qubit, highly
correlated with the primary qubit (which is subjected to the unknown
process), measurements taken in coincidence on the output 2-qubit state
allow reconstruction of the single-qubit process (this method is referred to as
ancilla-assisted process tomography -- AAPT).  Because this technique
requires correlation between the primary and ancillary qubits, maximally
entangled states yield the most accurate AAPT results (this special case of
AAPT is referred to as entanglement-assisted process tomography -- EAPT).  
Surprisingly, there is a class of separable -- completely unentangled -- states which possess the necessary correlations to perform AAPT \cite{D'Ariano4, aapt-prl}.  
All three types of QPT have been experimentally realized: SQPT, first in NMR \cite{qptresults, qptresults2} and later in photon systems~\cite{qptresults3}; EAPT~\cite{qptresults4, qptresults5}; and AAPT using unentangled states~\cite{aapt-prl}. 
Shown in Fig.~\ref{J3} are experimental arrangements for these three techniques as demonstrated in~\cite{aapt-prl}.

Experimentally, we have realized a variety of processes, allowing us
considerable freedom to manipulate our arbitrarily generated single-qubit
states.  Arbitrary unitary transformations can be generated with a series of
birefringent waveplates, 
specifically,  two quarter waveplates and a half waveplate~\cite{arbitrary-unitary}.  Incoherent full or partial projection is 
accomplished by using polarizers, which are inserted for a fraction of the 
total measurement time commensurate with the strength of the desired 
projection. Coherent partial projection is generated using tilted glass 
plates.  Perhaps the most interesting process, given current interest in 
quantum computation, is decoherence. As described in Sec. 2 and Appendix A, 
we introduce decoherence by passing photons through thick pieces of 
birefringent quartz.  By varying the thickness of quartz, an arbitrary 
strength of decoherence can be introduced.  By adding unitary transformations 
(waveplates) before and after the decoherer, this decoherence can be applied 
in any basis.

\begin{figure}[h!]
\begin{center}
\epsfig{file=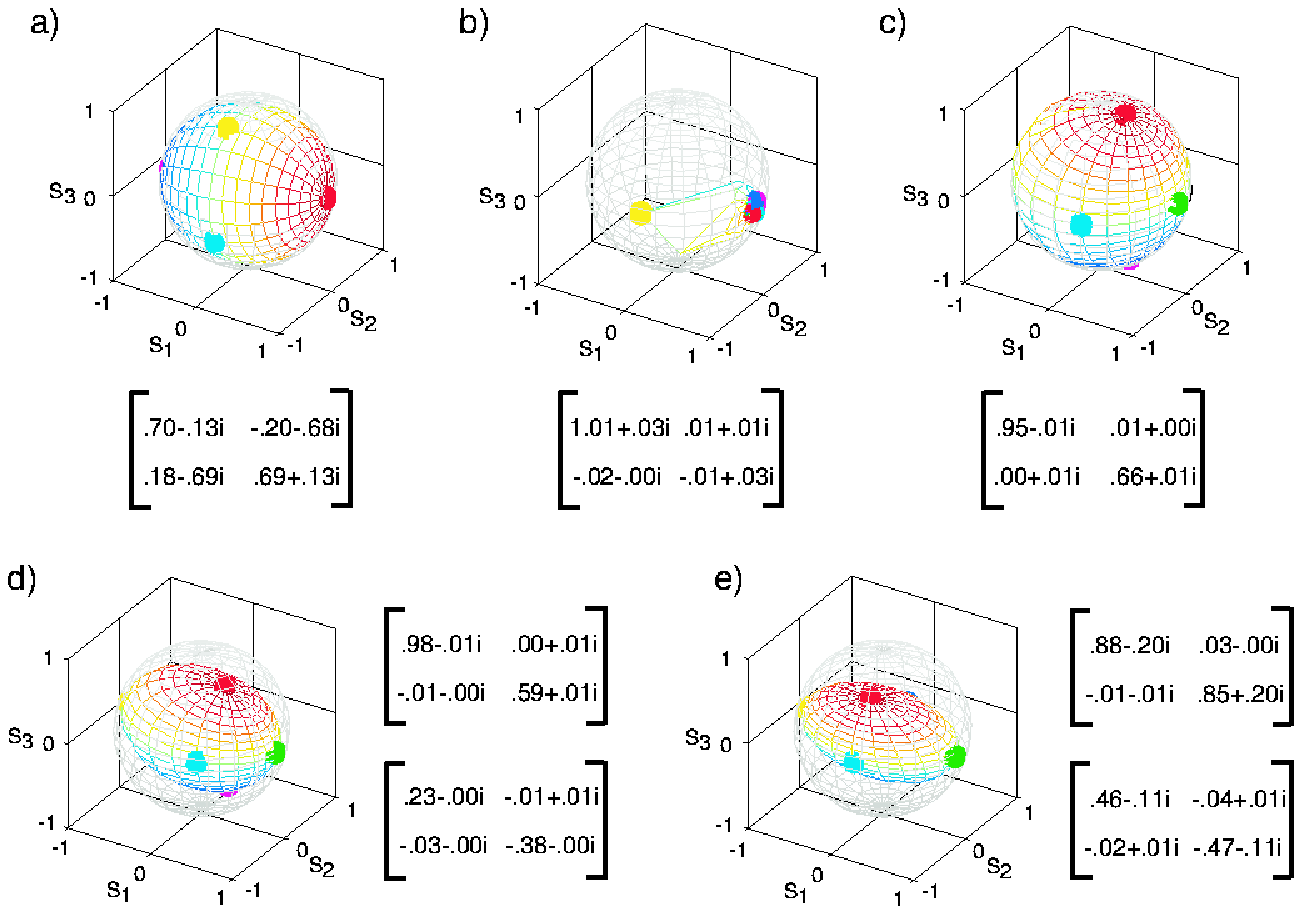, width=14cm}
\vspace{5 mm}
\fcaption{
Examples of experimentally generated and characterized processes are shown.  The
dark colored mesh corresponds to the output mapping of all input pure states.
The small solid spheres correspond to the cardinal points: H (green), 
V(yellow), D(purple), A(blue), R(red), and L(pink).  The 2x2 operator 
matrices correspond to the $E_j$ matrices which have a
greater than 1\% contribution to the measured process.  Note that in contrast to the operators in 
Fig. 6, here we have included the probability weightings directly in the 
matrices shown.  (a)  A unitary transformation implemented using a birefringent waveplate.
(b)  A horizontal polarizer.  Note that a single solid cardinal sphere is not
mapped to H.  This sphere corresponds to the vertical input case.  Some small
amount of vertical polarization leaks through this measured horizontal
polarizer.  Due to the extremely low number of vertical photons which survive
this process, the experimentally calculated output state is essentially
random.  This highlights one disadvantage in using this geometric picture:
intensity information is not directly visible.
(c)  A coherent partial polarizer.  Implemented using tilted glass slides,
this polarizer maintains the coherence of output states:  thus, pure states
remain pure, but slide along the surface of the sphere towards H.
(d)  An incoherent partial polarizer.  This was simulated by inserting a
horizontal polarizer into the beam 50\% of the time.
(e)  A decoherer.  A thick piece of birefringent quartz causes the H and V
polarizations to separate from each other in time.  They then become
partially distinguishable, and pure superpositions of H and V become mixed.}
\label{J4}
\end{center}
\end{figure}

Experimentally measured processes corresponding to the theoretical examples
given before are shown in Fig.~\ref{J4}.  As before, the geometric picture
offers a convenient way to visualize an otherwise complex operation.  In
fact, considering that an arbitrary process includes 15 independent
parameters (12 to define four orthogonal 2x2 operators and 3 to assign
probabilities to them), the ability to visualize its operation at all is
somewhat surprising.

The natural extension of this work is toward control of multiple qubits.  As
this work shows, in 
these exponentially larger Hilbert spaces 
it will be increasingly important to find
geometric or otherwise intuitive methods to solve the much harder problems
of multiple qubit creation, characterization, and manipulation.

\section{Acknowledgments}
We would like to thank Tzu-Chieh Wei and Daniel James for helpful
discussions, and acknowledge financial support from 
the DCI Postdoctoral Research Fellowship
Program, ARDA, and the National Science Foundation (Grant \#EIA-0121568).

\appendix
\noindent
Decoherence always arises from the entanglement of the quantum system
being considered to some other quantum degree of freedom, which is
then traced over.  In our system we realize decoherence by coupling
the frequency and polarization degrees of freedom and then measuring
in a frequency-insensitive way~\cite{berglund}.  The state of a
photon written in terms of its polarization and frequency spectrum
is $|\xi\rangle=(\alpha|H\rangle+\beta|V\rangle)\otimes\int d\omega
A(\omega)|\omega>$, where $\alpha$ and $\beta$ are complex normalized
coefficients, and $A(\omega)$ is the complex amplitude for frequency
$\omega$, normalized such that $\int d\omega|A(\omega)|^2=1$.  

To decohere in the $|H\rangle$/$|V\rangle$ basis, we send the
photon through a birefringent element whose fast axis is parallel
to the horizontal polarization.  Traversing a birefringent
element of length L adds a phase that is polarization and
frequency dependent, producing the state $|\xi_D\rangle =\int
d\omega (e^{i\frac{n_H L}{c}\omega}\alpha|H\rangle+e^{i\frac{n_V
L}{c}\omega}\beta|V\rangle)A(\omega)|\omega>$.	Tracing over the frequency
gives 
\begin{equation}
|\xi_D\rangle \rightarrow \rho_D = \bordermatrix{& \cr
&|\alpha|^2 &\alpha \beta^\ast \int_{-\infty}^{\infty} d\omega
|A(\omega)|^2e^{i \phi (\omega )} \cr
&\alpha^\ast \beta \int_{-\infty}^{\infty} d\omega |A(\omega)|^2e^{-i \phi
(\omega )} &|\beta|^2 \cr}, 
\label{decoh}
\end{equation}
where $\phi(\omega)=(n_H-n_V)\frac{L}{c}\omega$. As long as $(n_H-n_V)L$ is
much greater than the photon's coherence length\footnote{The coherence
length ($L_c$) of the photon is determined by the Fourier transform of the spectrum $A(\omega)$.
For example, if $A(\omega)$ is a Gaussian of full width at half maximum
$\Delta \omega$, then $L_c=\frac{2 \pi c}{\Delta\omega}$.}, the off-diagonal
elements of expression (\ref{decoh}) will effectively average to zero,
and the polarization state will be fully decohered.  Note that although
this form of decoherence due to dephasing is reversible (i.e., by using
a compensating birefringent element), it is fundamentially no different
than any other type of decoherence, which {\it in principle} is always
reversible if one could access the entire Hilbert space describing all
parts of the experiment.

\end{document}